# Socio-spatial Inequalities in a Context of "Great Economic Wealth"

## *Case study of neighbourhoods of Luxembourg City*

Natalia Zdanowska[1, 2, 3]

[1] Luxembourg Institute of Socio-Economic Research, Esch-sur-Alzette, Luxembourg
[2] Centre for Advanced Spatial Analysis, University College London, London, United Kingdom
[3] UMR CNRS 8504, Géographie-cités, Paris, France
natalia.zdanowska@liser.lu

## Abstract

In spite of being one of the smallest and wealthiest countries in the European Union in terms of GDP per capita, Luxembourg is facing socio-economic challenges due to recent rapid urban transformations. This article contributes by approaching this phenomenon at the most granular and rarely analysed geographical level – the neighbourhoods of the capital, Luxembourg City. Based on collected empirical data covering various socio-demographic dimensions for 2020–2021, an ascending hierarchical classification on principal components is set out to establish neighbourhoods' socio-spatial patterns. In addition, $Chi^2$ tests are carried out to examine residents' socio-demographic characteristics and determine income inequalities in neighbourhoods. The results reveal a clear socio-spatial divide along a north-west/south-east axis. Moreover, classical factors such as gender or citizenship differences are revealed to be poorly determinant of income inequalities compared with the proportion of social benefits recipients and single residents.

## Introduction

Luxembourg City has been undergoing radical urban transformations in the two last decades: a twofold population increase, a 75% rise in the number of foreign residents, who now represent more than 70% of the overall population, coupled with a 35% growth in employment. All these changes testify to the strong economic attraction of the capital of Luxembourg – one of the wealthiest countries in the European Union in terms of GDP per capita and minimum gross wage.[1] However, even if inequality is considered lower in wealthy nations, it is the richest countries that have experienced the strongest inequality growth over the past thirty years, especially in cities (OECD, 2016; Cottineau and Pumain, 2022). In fact, the GDP indicator constructed at national level is criticised for not reflecting many aspects of society and the well-being of residents.

---

[1] The minimum gross wage was equal to €2141 per month in 2020 (for unqualified employees), which is very high for the European Union (compared, for example, with €1539 in France or only €187 in Bulgaria) (Eurostat, 2023).

In the case of Luxembourg, even if the average gross monthly income[2] in 2021 in Luxembourg City was equal to €5714 and half of the residents received more than €4203 (gross per month), the largest percentage of residents with an income have an average gross monthly income of between €2300 and €3000 and almost 6% of all residents receive less than €1400 (Zdanowska, 2023). Moreover at the level of the municipality, the 20% of people with the highest income in Luxembourg City earn, on average, 8 times more than the 20% with the lowest, and this gap is greatest in particular neighbourhoods (Zdanowska, 2023). Indeed, at a more granular level, Luxembourg's rapid transformations have exacerbated socio-spatial inequalities and revealed challenges that need to be addressed by local authorities, such as land pressure, integration of foreign residents and social and territorial cohesion. This context involves relying on the concepts of social and spatial justice and desirable societal development (Rawls, 1971) in order to investigate cities experiencing an economic and demographic boom and the consequent inequalities.

In social science studies, class and social strata are essential factors in explaining society's various dynamics and phenomena regarding inequality, from classical discussions by Marx (1977) to Piketty (2020) and other studies in recent years (Pekkanen et al., 1995; Chan and Goldthorpe, 2007; Kingstone, 2000; Wright, 2005; Grusky, 2014). Social inequality refers to a state in which factors affecting human activities such as opportunities, resources (health, education, occupation, housing), and power, are unfairly distributed (Sen, 1992) – the last resulting in disparities across gender, race/ethnicity, class and other important social markers. Socio-spatial inequality then occurs in a situation of construction of spatial patterns or spatially based homogeneous groupings and is not evenly distributed across different geographical locations, which are constant over time (Han, 2022).

Most studies examining the distribution of social inequalities in urban space rely on income data and only capture economic or wealth inequalities (Pfeffer and Waitkus, 2021) or only one dimension of social inequality at a time, such as gender or age (Rashid, 2016). It is widely acknowledged, however, that inequality is a multifaceted phenomenon (Sen, 1992). Some studies also take into account other socio-economic aspects, such as occupation, education (Jung et al., 2014; Kernan and Bruce, 1972; Henning and Liao, 2013), public services or life expectancy (Lee and Rodríguez-Pose, 2013; Panori and Psycharis, 2017). Other multidimensional analyses of space provide a different perspective on its socioeconomic structure (Hacker et al., 2013; Lelo et al., 2019; Lin et al., 2015; Nijman and Wei, 2020; Spector, 1982; Zambon et al., 2017). This article privileges the income component, but also takes into account other socio-demographic factors to characterise socio-spatial inequalities. In Luxembourg, the latter have mostly been analysed at the national or municipality level. Neighbourhood analysis remains scarce and irregularly carried out over time (FOREG, 2008 ; Durand and Zdanowska, 2022 ; Zdanowska and Durand, 2023; Zdanowska, 2023).

This study will address the following questions: where are social inequalities located in the city and what patterns and similarities can be observed across neighbourhoods? What socio-demographic factors can explain these inequalities?

With this in view, a unique empirical dataset will be explored at the level of the 24 city-neighbourhoods, gathered from various social actors for 2020–2021 and covering 8 different dimensions: demography, migration, education, housing, employment, income and social benefits, delinquency. In a first step, two major statistical methods for classifying neighbourhoods in terms of socio-spatial patterns and similarities are applied: a principal component analysis (PCA) and an ascending hierarchical classification (AHC). Then, in a second step, contributory socio-demographic factors of income inequalities within neighbourhoods are tested by means of Chi² analysis.

---

[2] Referring in this article to wages and social benefits received by individuals. No other sources of income (capital) are registered in statistics in Luxembourg at neighbourhood level (Zdanowska, 2023).

## MATERIALS AND METHODS

The empirical data[3] consists of more than 100 indicators constructed and collected for 2020–2021 from various municipal departments of the City of Luxembourg and other national institutions that have never been simultaneously considered and have rarely been analysed at the level of the 24 administrative neighbourhoods (Durand and Zdanowska, 2022). Demographic and migration characteristics of residents were extracted from the municipal registry of the population (Registre de la population, Bierger Center). The residents' socio-demographic and socio-economic characteristics (jobs and income) were accessed via the GDPR compliant secure micro-data platform of the General Inspectorate of Social Security of the Luxembourg government (IGSS).[4] Data on seniors, housing, finances, education, youth and social intervention were provided by various municipal departments (Service Seniors, Service Logement, Service Finances, Service Enseignement, Service Jeunesse et Intervention Sociale) and volunteer organisations (such as Spëndchen asbl). Finally, data on municipal social benefits and crime were collected from the Social Office (Office Social) and the National Police Department, respectively.

Preparation and selection of the data was carried out to eliminate statistical biases and make comparisons between neighbourhoods possible. First, all the variables were expressed in relative values and standardised to eliminate the population size and extreme values effect. Second, a selection of significant and non-redundant variables for the study was made, by applying autocorrelation tests. Thus, from the initial 100 variables collected, 29 variables were selected (see Appendix 1) to run the PCA.

A PCA is an extremely powerful tool for compressing data information by reducing the number of variables to thematic main components or factorial axes, summarising several variables (Béguin and Pumain, 2000). A PCA enables the analysis of oppositions and similarities of neighbourhoods between the most relevant components, and reveals common neighbourhood characteristics, all dimensions of social inequality being considered simultaneously. Three axes provided a sufficiently significant rate (54% of a variance explained) for interpretation and the PCA therefore allowed a reduction of the 29 variables to 3 main components.

To obtain the final typology, a second reduction of the information was conducted by applying an AHC to the PCA results. An AHC is a statistical technique aimed at grouping entities – here neighbourhoods – so that the entities are as similar as possible within the same class (intra-class homogeneity maximisation), and the classes present the most dissimilar characteristics between them (maximising inter-class heterogeneity). The AHC results highlight 8 distinct classes, which come in two forms: 5 clusters, bringing together neighbourhoods with similar characteristics; 3 unique neighbourhood profiles with very distinct and unique aspects. A final typology of three main "groups" of neighbourhoods was proposed, after a validation of results by policy makers most aware of the city's social context.

---

[3] This work was carried out within the Social Observatory of the City of Luxembourg project financed by the City of Luxembourg for 2021–2024.
[4] The methodological limits in relation to the data source are the following: lack of information at the level of households in Luxembourg, only gross incomes available, other sources of income not available such as those from capital or those of international (NATO) and European civil servants (around 14,000 residents).

Finally, a last step consisted in examining the relationships between levels of income and socio-demographic factors in neighbourhoods. Bivariate non-parametric Chi² tests were applied as the majority of variables accessible within neighbourhoods were qualitative, such as gender, citizenship, NACE activity sector codes or civil status of residents. The only quantitative variable (individual income) was transformed into a qualitative variable and divided into income classes. In total, 11 variables (see Appendix 2) were selected for the Chi² tests. Thus 220 Chi² tests were carried out across all variables for all neighbourhoods.[5]

## RESULTS

### *The socio-economic and socio-demographic divide*

The first PCA axis (Appendix 3) explains most of the similarities and differences between neighbourhoods (29% of variance). This component indicates a clear socio-economic opposition. On the one hand, the more "wealthy" neighbourhoods (in blue) are characterised by a high median income, an important share of well-paid jobs in the public sector, a significant inter-district income inequality and a relatively large share of jobs in the specialist occupations in the scientific and technical sectors. On the other hand, in the "less well-off" neighbourhoods (in red), the proportion of recipients of social benefits, such as the Social Inclusion Income (REVIS), the municipality's solidarity allowance and unemployment benefit, is higher. The proportion of residents registered at the Social Office and the proportion of social housing in relation to the population of the district is also high.

The second PCA axis (Appendix 4), concentrating on 15% of the initial information, is characterised by both demographic and social components, bringing information on the structure of the population (age and nationality), but also on residential mobility. Less well-off neighbourhoods (in red) with strong residential turnover and a substantial proportion of people living below the poverty line (with income below 60% of the national median) contrast with more well-off neighbourhoods (in blue), populated by Luxembourgish citizens (relatively more compared with other neighbourhoods), with a higher share of employment in the public sector than the city average, and with a relatively high average age.

Finally, the third axis (Appendix 5) explains 12% of all the information and is mostly determined by, on the one hand, neighbourhoods with high residential turnover (in blue) and, on the other, those with high employment rates (in red). This axis is also determined by the "education" component, which differentiates neighbourhoods where the proportion of children enrolled in private schools compared with all children in school is substantial, and neighbourhoods where the proportion of

---

[5] For confidentiality reasons small neighbourhoods had to be merged (Clausen with Pfaffenthal and Hamm with Pulvermuhle), making 22 neighbourhoods out of the 24 initially analysed.

Luxembourgish children enrolled in public schools is significant (compared with other nationalities), bearing in mind that Luxembourgish children are a minority in the city.

### *The "north-west and north-east" vs the "south-east and north-south"*

The final socio-spatial typology of the neighbourhoods of the capital city produced by an AHC on PCA results made it possible to indicate three socially distinct groups of neighbourhoods. The first type are those located in the north-western and north-eastern parts of the city (see Figure 1). They are characterised by a very high median salary, a high proportion of jobs in the public sector and high-skilled sectors (specialist occupations in scientific and technical activities, finance and insurance). These "well-off" neighbourhoods are, in parallel, also marked by income inequalities within neighbourhoods, which means that, despite the relatively high level profiles of residents, they are not homogenous. However, significant differences should be pointed out between neighbourhoods within this large group.

In Belair, Limpertsberg, Kirchberg, Merl, Neudorf, Rollingergrund and Weimerskirch (in blue, Figure 1), cosmopolitanism is marked, the residents are among the best paid in the capital, the average age is lower than in other neighbourhoods, and income inequalities within neighbourhoods are significant. In Ville-Haute (in grey), the average age of residents is very high, which can be explained by the presence there of the main retirement residential developments (including the biggest one Fondation Pescatore). The residents in Ville-Haute are rather Luxembourgish, with high incomes and very homogeneous socio-economic profiles (little difference in income levels). The proportion of those in receipt of long-term care is high and residential rotation is very significant. Finally, residents of Cents (in green) are mainly Luxembourgish, and the proportions of employees in the public sector is relatively high. Cents is also a very homogeneous district in terms of socio-economic income profiles, which are among the highest in the city (but lower than in Ville-Haute). The residential turnover is, however, very low.

Neighbourhoods of the second type are located in the south-eastern part of the city and along a north-south axis (Figure 2), which also follows a historical division of the city outside the walls of the fortress, where the railways station was constructed and where the lower zones of the valley are located. The common socio-economic characteristics of these neighbourhoods are the following: median wages below the municipal average, very significant income inequalities within neighbourhoods, a high proportion of recipients of REVIS, unemployment benefit and the municipality's solidarity allowance, and of residents registered at the Social Office. The proportion of residents living below the poverty line is very high. Several clusters are part of this second major type of neighbourhoods.

First, the distinctive characteristic of Gare (in pink) compared with other neighbourhoods is that it has highest residential mobility rate in the city. In fact, Luxembourg City itself is, on the national scale, characterised by strong residential mobility. In 2020, more than 40,000 people moved house: including over 16,000 new residents, 14,800 who left the capital (half of whom remained in the country), and nearly 10,000 who moved but remained in the city. This indicator highlights the strong turnover of the population experienced by the city and raises questions about the reasons for this migratory dynamism, but also about the problems related to the integration of certain populations (Durand and Zdanowska, 2022). Moreover, Gare presents very significant income inequalities within the neighbourhood and income below the municipal average. It has a large proportion of social assistance recipients, a relatively younger population than other neighbourhoods and a much larger proportion of foreigners than elsewhere. Bonnevoie-Sud and Bonnevoie-Nord (in beige) are neighbourhoods with a low average age, income inequalities higher than the municipal average (but to a lesser extent than in Gare) and a relatively high number of social assistance recipients. These two neighbourhoods are also characterised by a large proportion of children enrolled in public schools and a high employment rate.

In Grund, Pulvermühle and Pfaffenthal – neighbourhoods located in the valley – and Hamm (in orange, Figure 1), the average age is high, the proportion of Luxembourgish citizens is substantial, the proportion of children enrolled in public schools is higher than the municipal average, and the number of social assistance recipients is high. Finally Clausen and Eich (historically a more industrial area) are neighbourhoods (in grey, Figure 1) where incoming and outgoing migratory flows are relatively high compared with other neighbourhoods in the city.

In a final third grouping, the rest of neighbourhoods are "intermediate" in the sense that their main common feature is that all the socio-demographic characteristics taken into account are close to the municipal average with no extremes values. It is then possible to formulate the hypothesis that this is where the "middle-class" of the Luxembourg City region is located.

***Social benefits recipients as markers of income inequalities – regardless of gender, age or nationality***

A final analysis aimed at examining the relationships between levels of income and socio-demographic factors in neighbourhoods by applying bivariate non-parametric $Chi^2$ tests. The results of the 220 $Chi^2$ tests highlight and empirically confirm a commonly accepted hypothesis: at the scale of the entire capital, among all the socio-demographic factors of the residents, the most determining factor of income inequalities in neighbourhoods is the greater or less significant presence of people receiving social benefits. Indeed, in Luxembourg City, the relationship between incomes and the number of people receiving the state cost of living allowance (COS) is the most important considering the Tschuprow's T ($T = 0.53$) (see Table 1). The link between income and the fact of being a REVIS recipient is the second most important relation ($T= 0.46$), followed by that between income and the fact of receiving unemployment benefit ($T = 0.25$). Thus the COS is a social benefit that is statistically more decisive than the REVIS for understanding the unequal distribution of incomes within a neighbourhood.

Moreover, the results point to an interesting observation: relationships between income and gender ($T=0.11$ for the municipality and $T=0.06$ in Clausen), age ($T= 0.17$), nationality ($T=0.14$), type of job contract ($T=0.14$), type of sector ($T=0.15$) and NACE sector ($T=0.22$), are very weak determinants of low income in a neighbourhood compared with social benefits, even if such a relationship does exist. Thus the fact of being in receipt of unemployment benefit, the REVIS and the COS is the strongest factor in determining precariousness in Luxembourg City, and greater than other factors that are often suggested such as nationality, age or gender.

In fact, in Bonnevoie Nord, Bonnevoie Sud, Gare, Clausen/Pfaffenthal and Eich, gender is the least strong determinant of low income (Tschuprow's T lesser than 0.10). The same conclusion can be drawn for the nationality factor (T=0.12), particularly in the case of Cessange (T=0.11) and Belair (T=0.10). This means that, in these neighbourhoods, the nationality of the residents (although of a very heterogeneous nature in the case of Belair) does not determine the differences in income that may exist in these neighbourhoods, despite its being a very commonly suggested reason.

A comparison between neighbourhoods highlights quite significant differences (Table 1). In Gare, the existing relationship between receipt of the COS and the unequal distribution of income can be characterised as the strongest (T = 0.76), followed by Eich (T=0.71). The REVIS is also a key factor in explaining income inequalities and this relationship is strongest in Rollingergrund (T= 0.68), Gare (T= 0.66) and Eich (T= 0.62). The type of employment contract (fixed-term or permanent) is ultimately not a very decisive determinant of income inequalities, although an explanation for this result lies in the fact that the majority of residents of Luxembourg City have a permanent contract.

The civil status of residents at the level of the municipality is an explanatory factor for unequal income distribution, but certainly less significant (T= 0.29) than the fact of being in receipt of social benefits such as REVIS or the COS, but it is just as important as, for example, receiving unemployment benefit (T= 0.30). It should also be noted that the results show that the relationship can be described as existing and moderate in all 22 neighbourhoods. Thus the civil status of a resident of a neighbourhood, and especially the fact of being single compared with being married or in a partnership, has an influence on income inequalities. This is most striking in the neighbourhoods of Belair (T= 0.43), Limpertsberg (T=0.41) and Ville-Haute (T=0.39), where the relationship is existing and strong. We may hypothesise that, even if these neighbourhoods are characterised by high gross average incomes, they have a high proportion of single residents, who, compared with married people or those in civil partnerships, have lower incomes. However, this interpretation should be compared with analyses at the household level.

The "unemployment benefit" factor it is the most determining in explaining income inequality in Gare (T= 0.40) and Bonnevoie-Sud (T=0.34), but also in Limpertsberg (T=0.34) and Merl (T=0.36). These results confirm that the previously identified so-called well-off neighbourhoods are also characterised by residents with different social statuses, and where income disparities are as significant as in Gare or Bonnevoie Sud, previously referred to as less "well-off" neighbourhoods. These results statistically confirm the hypothesis that, despite being the capital of the wealthiest country in the European Union, Luxembourg City presents a great diversity of income levels within different types of neighbourhoods and important spatial divisions between neighbourhoods.

# CONCLUSION

On the one hand, neighbourhoods of the first major type in the City of Luxembourg in terms of common socio-spatial trends are located in the north-western and north-eastern parts of the capital. The "north-west" axis brings together Belair, Kirchberg, Limpertsberg, Merl, Neudorf, Rollingergrund and Weimerskirch. These neighbourhoods are characterised by a pronounced multiculturalism and are, on average, populated by young residents of working age. The proportion of employment in highly-skilled sectors as (specialist scientific and technical activities, finance and insurance) is among the highest in the municipality. Incomes in these neighbourhoods are among the highest in the capital, but are at the same time very unevenly distributed within the neighbourhoods themselves. The proportion of students enrolled in private schools is also very high compared with the rest of the city.

The "north-east" axis of the city is made up of Ville-Haute and Cents which are, like the previous districts, among the most "well-off" in the city. The average age is very high and the population much more homogeneous in terms of nationality (Luxembourgish more represented than elsewhere) and income (high and with little disparity within the districts of Cents). In addition, the inhabitants of these two neighbourhoods are relatively less involved in social benefits, even though in Ville-Haute the proportion of recipients of long-term care is high, as is residential turnover, although this latter index is low in Cents.

On the other hand, neighbourhoods of the second major type on the "south-east" axis (Gare, Bonnevoie-Sud and Bonnevoie-Nord) present very atypical profiles. They are located on the other side of the historical fortress of Luxembourg City and developed after the construction of the railway station in the Gare district. The sociodemographic situation in this second group is relatively the most disadvantaged in the capital. Median incomes are the lowest with very significant wage disparities. The population is younger and more international than in other districts, and the proportion of employees living below the poverty line is the highest in the capital. The proportion of social assistance recipients is high. Within these neighbourhoods, there is a certain degree of social mix since residents with varied social profiles live side by side. Gare presents a very remarkable and much greater residential mobility than Bonnevoie-Nord or Bonnevoie-Sud, and has a hub function in the city.

The neighbourhoods of the "north-south" axis complete this second group of less well-off districts: Grund, Pulvermühle, Hamm, Pfaffenthal, Eich and Clausen. Located in the lower part of the city in the valley, which was historically poorer, they are in a rather unfavourable social position at the city level, but to a lesser extent than those on the "south-east" axis. They present a high average age, low population density and constitute a substantial proportion of the population. The average salary is generally lower than the municipal average, the median income is low compared with other neighbourhoods and strong income disparities within the neighbourhoods can be observed. The proportion of people receiving social assistance is generally very high compared with other neighbourhoods. The districts of Eich and Clausen stand out for their high residential turnover index.

In a geographical area close to the "north-west" axis, another group of so-called "intermediate" neighbourhoods (Beggen, Cessange, Dommeldange, Gasperich, Hollerich and Mühlenbach) presents an average profile, in the sense that the all the characteristics considered in the study are close to the municipal average.

Finally, complementary statistical analyses at the level of the city and the neighbourhoods highlighted that the socio-demographic variables commonly interpreted as having an influence on income inequalities (sex, age, nationality) are not so revealing and highly significant at our scale of analysis and given the number of statistical observations. Indeed, the most determining socio-demographic factors are linked to the precarious social situation of residents, expressed by their being in receipt of social benefits (unemployment benefit, REVIS or cost-of-living allowance). Indeed, the cost-of-living allowance is statistically the most determining factor for unequal income distribution. However, one demographic characteristic – marital status (being married, single or divorced) – emerged from the analysis as important in explaining income inequalities in a neighbourhood. This can be linked to the possible difficulties faced by single-parent families living in the capital.

To go further in the statistical analysis of the factors explaining income inequalities, it would be essential in future to be able to undertake analysis on a scale smaller than the neighbourhood (urban blocks or grid) and to be able to carry out analysis of quantitative variables between them. However, access to this type of information is a major challenge for the entire scientific community.

These various results certainly provide information on the composition and socio-economic characteristics of the population. Above all, however, they reveal significant spatial disparities and social discrepancies between residents. Many inhabitants of the capital find themselves in a precarious situation – in terms of food, employment and housing – and require support or social assistance.

The twofold increase in prices in the past twenty years, the pressure on the housing market (Mezaros and Paccoud, 2022) and the insufficient amount of affordable and social housing raise the issue of maintaining social mix and offering possibilities for disadvantaged populations to stay in the city. These findings open up a discussion on the difficulties encountered by individuals willing to live and stay in the capital city, and the challenges faced by the public authorities in meeting their expectations in order to achieve more spatial justice.

**Figure 1 First type of neighbourhood (AHC on PCA)**

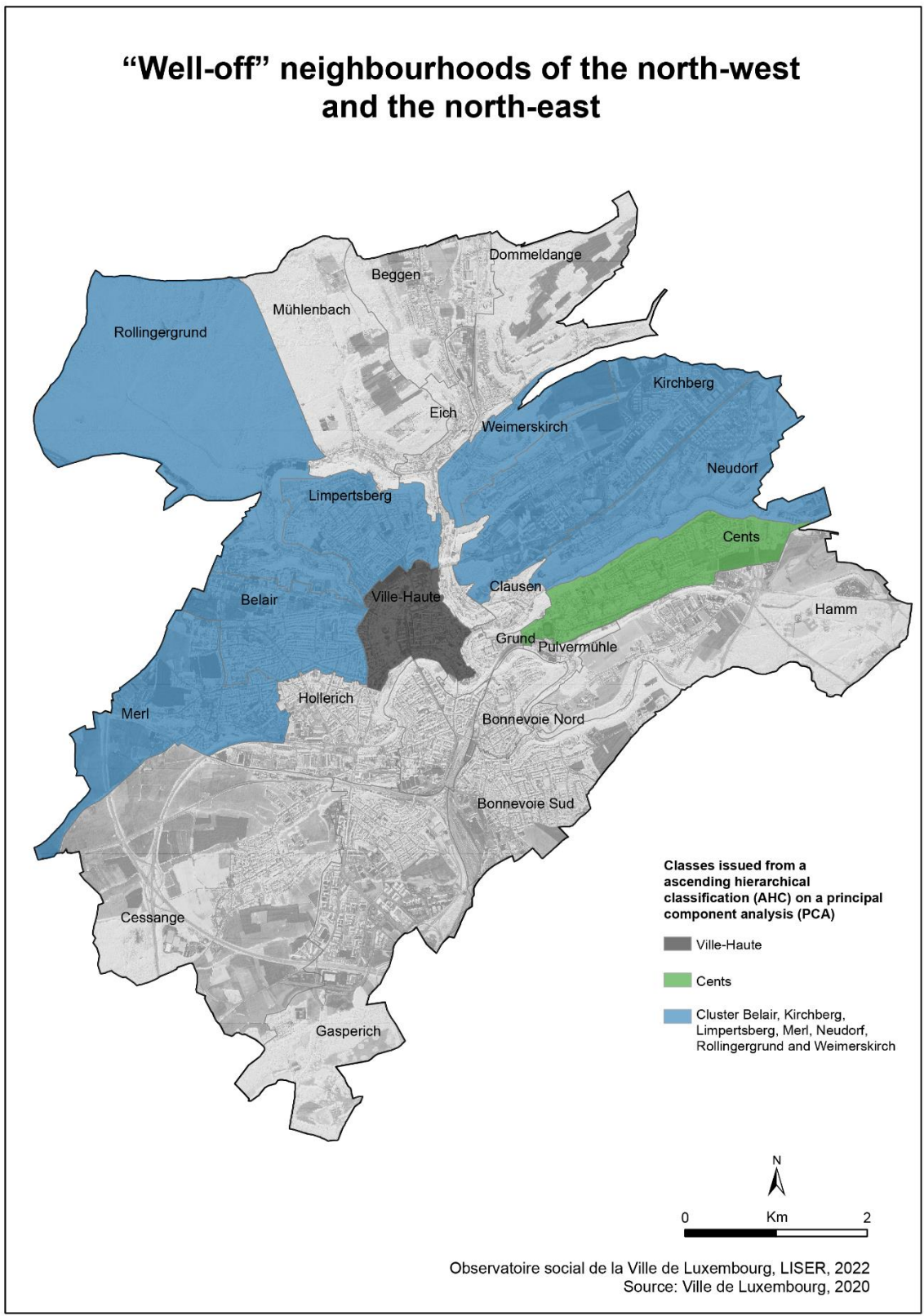

**Figure 2 Second type of neighbourhood (AHC on PCA)**

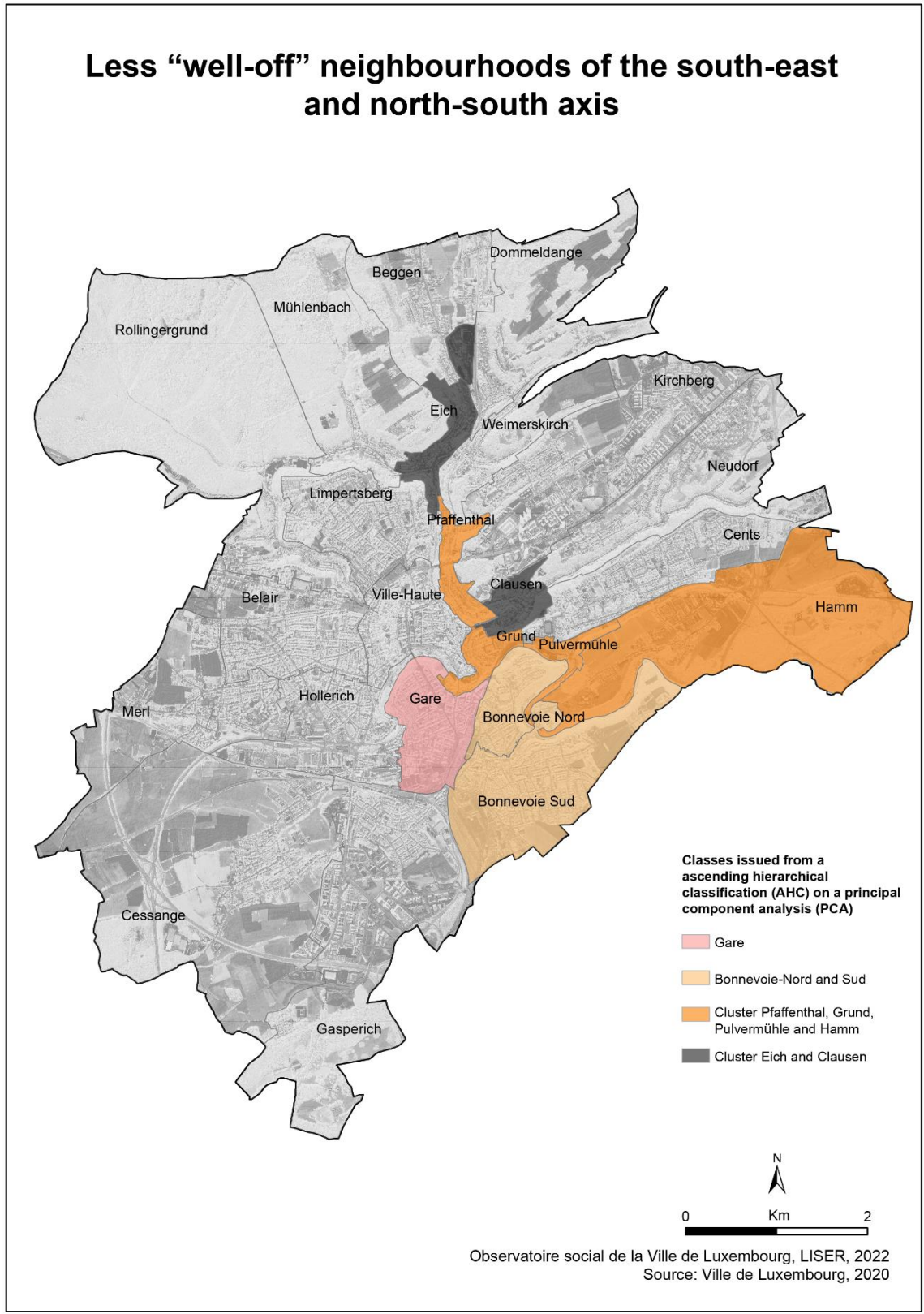

**Table 1 Chi² test - value of Tschuprow's T by neighborhood for each pair of relationship between income and ten socio-demographic variables considered**

| T de Tschuprow | Gender | Age | Status | Citizenship | Contract type | Sector type | Sector (NACE) | Unemployment | REVIS | COS |
|---|---|---|---|---|---|---|---|---|---|---|
| Beggen | 0,13 | 0,18 | 0,25 | 0,17 | 0,17 | 0,16 | 0,22 | 0,17 | 0,30 | 0,35 |
| Belair | 0,10 | 0,16 | 0,43 | 0,10 | 0,10 | 0,18 | 0,16 | 0,32 | 0,46 | 0,57 |
| Bonnevoie Nord | 0,08 | 0,16 | 0,28 | 0,15 | 0,14 | 0,18 | 0,19 | 0,30 | 0,41 | 0,55 |
| Bonnevoie Sud | 0,07 | 0,12 | 0,26 | 0,13 | 0,11 | 0,16 | 0,18 | 0,34 | 0,60 | 0,64 |
| Cents | 0,13 | 0,14 | 0,27 | 0,12 | 0,12 | 0,15 | 0,16 | 0,30 | 0,34 | 0,51 |
| Cessange | 0,10 | 0,15 | 0,24 | 0,11 | 0,15 | 0,16 | 0,19 | 0,33 | 0,32 | 0,36 |
| Clausen/Pfaffenthal | 0,06 | 0,17 | 0,24 | 0,16 | 0,16 | 0,16 | 0,22 | 0,25 | 0,45 | 0,41 |
| Dommeldange | 0,07 | 0,15 | 0,26 | 0,14 | 0,13 | 0,16 | 0,20 | 0,28 | 0,46 | 0,45 |
| Eich | 0,06 | 0,15 | 0,26 | 0,13 | 0,18 | 0,11 | 0,18 | 0,32 | 0,62 | 0,71 |
| Gare | 0,05 | 0,14 | 0,35 | 0,14 | 0,13 | 0,12 | 0,18 | 0,40 | 0,66 | 0,76 |
| Gasperich | 0,09 | 0,16 | 0,32 | 0,14 | 0,14 | 0,13 | 0,17 | 0,33 | 0,46 | 0,55 |
| Grund | 0,16 | 0,19 | 0,30 | 0,18 | 0,13 | 0,20 | 0,26 | 0,16 | 0,34 | 0,36 |
| Hamm/Pulvermühle | 0,09 | 0,16 | 0,25 | 0,17 | 0,13 | 0,20 | 0,21 | 0,27 | 0,41 | 0,41 |
| Hollerich | 0,08 | 0,15 | 0,34 | 0,13 | 0,13 | 0,14 | 0,17 | 0,33 | 0,51 | 0,62 |
| Kirchberg | 0,12 | 0,13 | 0,24 | 0,12 | 0,12 | 0,13 | 0,18 | 0,27 | 0,45 | 0,63 |
| Limpertsberg | 0,10 | 0,15 | 0,41 | 0,10 | 0,13 | 0,16 | 0,17 | 0,34 | 0,50 | 0,60 |
| Merl | 0,08 | 0,16 | 0,27 | 0,12 | 0,11 | 0,15 | 0,17 | 0,36 | 0,46 | 0,51 |
| Mühlenbach | 0,08 | 0,16 | 0,20 | 0,15 | 0,14 | 0,15 | 0,19 | 0,24 | 0,43 | 0,46 |
| Neudorf | 0,09 | 0,14 | 0,28 | 0,12 | 0,16 | 0,15 | 0,18 | 0,32 | 0,32 | 0,53 |
| Rollingergrund | 0,07 | 0,16 | 0,28 | 0,14 | 0,17 | 0,15 | 0,17 | 0,26 | 0,68 | 0,62 |
| Ville-Haute | 0,12 | 0,18 | 0,39 | 0,13 | 0,09 | 0,19 | 0,22 | 0,22 | 0,42 | 0,53 |
| Weimerskirch | 0,11 | 0,17 | 0,23 | 0,14 | 0,14 | 0,15 | 0,18 | 0,25 | 0,44 | 0,52 |
| **Luxembourg-ville** | 0,09 | 0,16 | 0,29 | 0,13 | 0,13 | 0,16 | 0,19 | 0,30 | 0,46 | 0,53 |

T < 0,10 Very weak or no relationship
0,10 < T < 0,20 Existing relationship
0,20 < T < 0,35 Moderate existing relationship
0,35 < T < 0,49 Strong existing relationship
T > 0,50 Very strong existing relationship

**Appendix 1 Quantitative socio-demographic variables (PCA and AHC)**

|  | Indicator | Source |
|---|---|---|
| 1 | Density | Registre de la population, Bierger center, 2020 |
| 2 | Majority of men over women | Registre de la population, Bierger center, 2020 |
| 3 | Average age | Registre de la population, Bierger center, 2020 |
| 4 | Residential turnover[6] | Registre de la population, Bierger center, 2020 |
| 5 | Proportion of recipients of "assurance dependence" social aid compared with the population of the neighbourhood | Inspection générale de la sécurité sociale, 2021 |
| 6 | Proportion of population aged over 65 years, recipient of the téléalarmes[7] service | Service Seniors, Ville de Luxembourg, 2020 |
| 7 | Proportion of Luxembourgish among the total population of the district | Registre de la population, Bierger center, 2020 |
| 8 | Proportion of social housing in relation to the total population of the district | Service logement + other organisations (Fonds du logement, SNHBM, AIS…), 2021 |
| 9 | Average selling price of housing per m² | Observatoire de l'Habitat, 2020-2021 |
| 10 | Average annual growth rate of selling prices of housing units per m² between 2009 and 2020 | Observatoire de l'Habitat, 2020-2021 |
| 11 | Employment rate | Inspection générale de la sécurité sociale, 2021 |
| 12 | Median salary | Inspection générale de la sécurité sociale, 2021 |
| 13 | Wage inequalities within neighbourhoods | Inspection générale de la sécurité sociale, 2021 |
| 14 | Wage inequalities between neighbourhoods compared with the municipal average | Inspection générale de la sécurité sociale, 2021 |
| 15 | Proportion of jobs in the finance and insurance sectors among all employed people | Inspection générale de la sécurité sociale, 2021 |
| 16 | Proportion of jobs in the sector of specialist scientific and technical activities among all employed persons | Inspection générale de la sécurité sociale, 2021 |
| 17 | Proportion of jobs in the public sector among all employed people | Inspection générale de la sécurité sociale, 2021 |
| 18 | Proportion of working-age population receiving REVIS | Inspection générale de la sécurité sociale, 2021 |
| 19 | Proportion of the population of working age receiving unemployment benefit | Inspection générale de la sécurité sociale, 2021 |
| 20 | Proportion of the population living below the wage insecurity threshold (with a salary below 60% of the median) | Inspection générale de la sécurité sociale, 2021 |
| 21 | Proportion of recipients of housing assistance among the total population of the district | Service Finances, Ville de Luxembourg, 2020 |
| 22 | Proportion of recipients of the municipality's solidarity allowance among the total population of the district | Service Jeunesse et Intervention Sociale, Ville de Luxembourg, 2020 |
| 23 | Proportion of the population registered in social grocery stores (épiceries sociales) among the total population of the district | Spendchen, Ville de Luxembourg, 2020 |
| 24 | Proportion of people registered with the Social Office (Office Social) among the total population of the district | Office social, Ville de Luxembourg, 2021 |
| 25 | Proportion of burglaries among the total population of the district | Police, Ville de Luxembourg, 2020 |
| 26 | Proportion of domestic violence per 1000 inhabitants | Police, Ville de Luxembourg, 2020 |
| 27 | Proportion of Luxembourgish pupils in public schools among all pupils in public schools | Service Enseignement, Ville de Luxembourg, 2021 |
| 28 | Proportion of Luxembourgish pupils in private schools among all pupils in private schools | Service Enseignement, Ville de Luxembourg, 2021 |
| 29 | Share of students enrolled in private schools among all schoolchildren | Service Enseignement, Ville de Luxembourg, 2021 |

---

[6] The residential turnover indicator measures the volume of residential movements, taking into account both incoming flows (i.e. linked to immigration, but also to internal migration into the capital) and outgoing flows (linked to emigration and intra-urban moves).

[7] The *téléalarme service* is a home alert system offered by the city for urgent medical assistance calls.

**Appendix 2 Qualitative socio-demographic variables and modalities (Chi² tests)**

|   | Qualitative variables | Modalities |
|---|---|---|
| 1 | Gender | Men; Women |
| 2 | Age | Less than 14 years; 14-19 years; 20-24 years; 25-29 years; 30-34 years; 35-39 years; 40-44 years; 45-49 years; 50-54 years; 55-59 years; 60-64 years; 65-69 years; 70-74 years; more than 75 years old |
| 3 | Civil status | Married, Single, Divorced |
| 4 | Citizenship | Luxemburg; Germany; Belgium; France; Portugal; other European Union (EU) nationality; nationality outside the EU |
| 5 | Type of employment contract | Permanent contract, fixed-term contract, apprenticeship contract |
| 6 | Activity sector | NACE Rev.2 classification (letters A-U) |
| 7 | Type of activity sector | Public, Private, Independent |
| 8 | Unemployment benefit | Recipient ; Non-recipient |
| 9 | Minimum wage (Revenu d'Inclusion Sociale – REVIS) | Recipient ; Non-recipient |
| 10 | Cost of Living Allowance (Allocation de vie chère) | Recipient ; Non-recipient |
| 11 | Gross monthly income | Less than €1400; €1401 – €2300; €2301 – €3000; €3001 – €3700; €3701 – €4500; €4501 – €5200; €5201 – €5900; €5901 – €6600; €6601 – €7300; €7301 – €8000; €8001 – €8800; €8801 – €9500; €9501 – €10200; €10201 – €11700; €11701 – €12400; €12401 – €18100; more than €18100. |

**Appendix 3 PCA Axis 1**

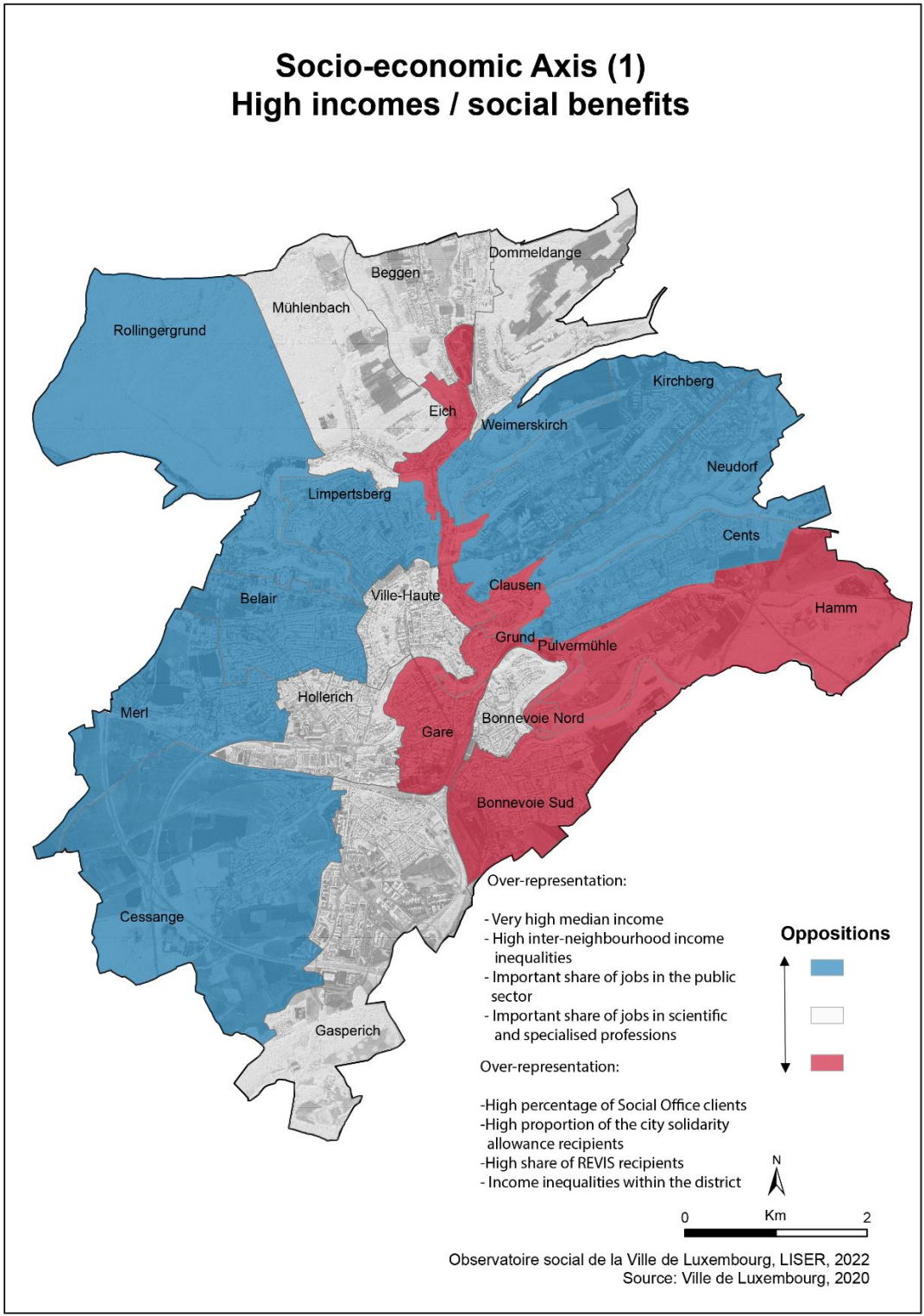

**Appendix 4 PCA Axis 2**

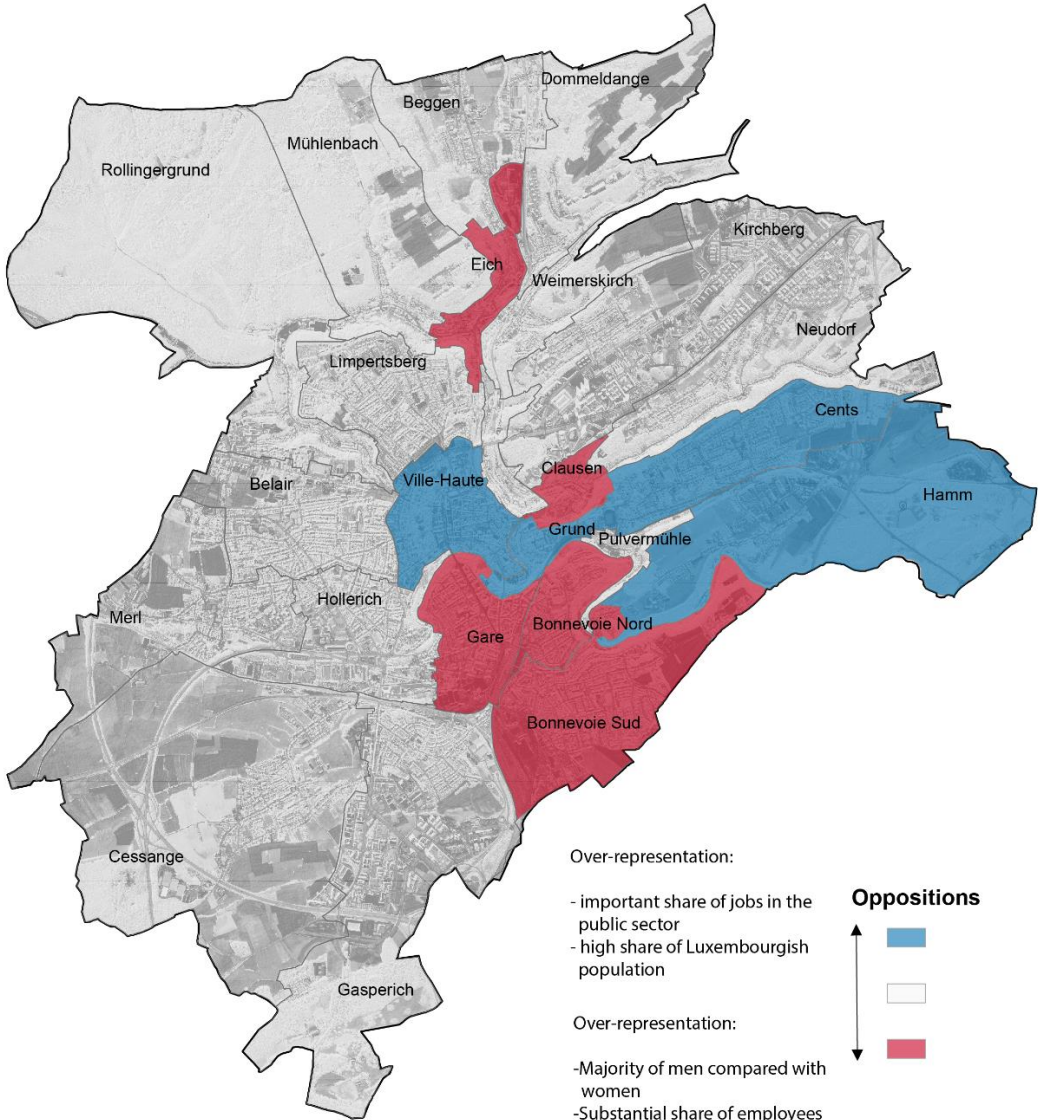

**Appendix 5 PCA Axis 3**

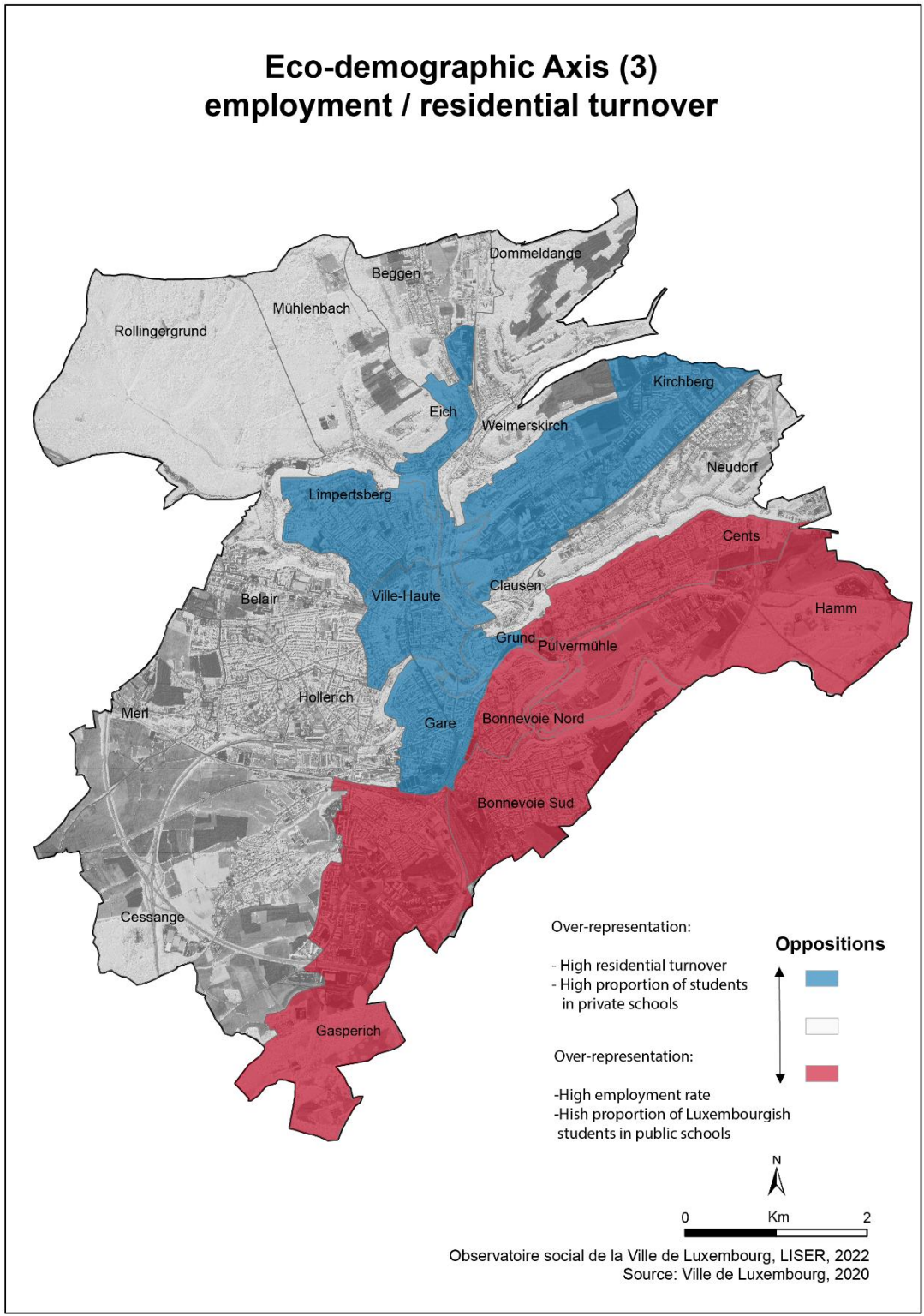